\documentclass[structabstract]{aa}

\usepackage{txfonts}          
\usepackage{mathrsfs}         
\usepackage{rotating}         
\usepackage{subfigure}        
\usepackage{lscape}           
\usepackage{afterpage}        
\usepackage{xspace}           
\usepackage{natbib}           
\usepackage{graphicx}
\usepackage{epstopdf}
\usepackage{indentfirst}
\usepackage{chemfig}
\usepackage{multicol}
\usepackage{booktabs,dcolumn}
\usepackage[T1]{fontenc}
\usepackage{float}

\title{Rotational spectroscopy of the HCCO and DCCO radicals in the millimeter and submillimeter range \thanks{The full list of transitions only available at the CDS via anonymous
ftp to cdsarc.u-strasbg.fr (130.79.128.5) or via http://cdsarc.u-strasbg.fr/viz-bin/qcat?J/A+A/vol/page}}

\author{J.~Chantzos\inst{1} \and S.~Spezzano\inst{1}  \and C.~Endres\inst{1} \and L. Bizzocchi\inst{1} \and V.~Lattanzi\inst{1} \and J.~Laas\inst{1} \and A.~Vasyunin\inst{2,3} \and P.~Caselli\inst{1}}

\institute{Center for Astrochemical Studies, 
           Max-Planck-Institut f\"ur extraterrestrische Physik,
           Gie\ss enbachstra\ss e 1, 85748 Garching (Germany) \and Ural   
           Federal University, Ekaterinburg, Russia \and Visiting Leading Researcher, Engineering Research Institute
`Ventspils International Radio Astronomy Centre' of Ventspils
University of Applied Sciences, In\v{z}enieru 101, Ventspils LV-3601,
Latvia}

\titlerunning{Rotational spectroscopy of the HCCO and DCCO radicals in the millimeter and submillimeter range}
\authorrunning{J. Chantzos et al.}

\begin{document}

\abstract
{The ketenyl radical, HCCO, has recently been detected in the ISM for the first time. Further astronomical detections of HCCO will help us understand its gas-grain chemistry, and subsequently revise the oxygen-bearing chemistry towards dark clouds. Moreover, its deuterated counterpart, DCCO, has never been observed in the ISM. HCCO and DCCO still lack a broad spectroscopic investigation, although they exhibit a significant astrophysical relevance.}
{In this work we aim to measure  the pure rotational spectra of the ground state of HCCO and DCCO in the millimeter and submillimeter region, considerably extending the frequency range covered by previous studies.}
{The spectral acquisition was performed using a frequency-modulation absorption spectrometer between 170 and 650 GHz. The radicals were produced in a low-density plasma generated from a select mixture of gaseous precursors. For each isotopologue we were able to detect and assign more than 100 rotational lines. }
{The new lines have significantly enhanced the previous data set allowing the determination of highly precise rotational and centrifugal distortion parameters. In our analysis we have taken into account the interaction between the ground electronic state and a low-lying excited state (Renner-Teller pair) which enables the prediction and assignment of rotational transitions with  $K_a$ up to 4. }
{The present set of spectroscopic parameters provides highly accurate, millimeter and submillimeter rest-frequencies of HCCO and DCCO for future astronomical observations. We also show that towards the pre-stellar core L1544, ketenyl peaks in the region where $c$-$\mathrm{C_3H_2}$ peaks, suggesting that HCCO follows a predominant hydrocarbon chemistry, as already proposed by recent gas-grain chemical models.}


\maketitle

\section{Introduction} \label{intro}
The evolution of organic molecules from dark clouds to planetary systems is  of fundamental importance to shed light on the astrochemical heritage of life on Earth \citep{case_cecca}. However, interstellar organic chemistry is still not fully understood \citep{herbst}. Complex organic species (hereafter COMs) have been widely detected in hot cores embedded in high-mass star forming regions \citep{cummins,blake}, and low-mass hot corinos \citep{cazaux,bottinelli}. Their formation towards warm sources is well reproduced by the so-called "warm-up scenario" \citep{garrod06,garrod08}, wherein COMs form on the surface of dust grains through radical-radical reactions and are eventually ejected to the gas-phase via thermal desorption during the warm-up phase of the protostar. Organic species, such as $\mathrm{CH_3OCH_3}$, $\mathrm{H_2CCO}$ and  $\mathrm{CH_3OCHO}$ have been detected also in cold, pre-stellar cores \citep{cernicharo,bacmann, vastel, jimenez} but need other formation processes to justify their presence in the gas phase, since there is no rise in temperature toward a starless or a pre-stellar core ($\mathrm{T < 20 \, K}$). Pre-stellar cores are the initial phase of star formation right before gravitational collapse sets in, which ultimately leads to the birth of a protostar. 

So far, there have been several chemical models \citep{vasyunin2013, reboussin, ruaud, chang} that try to reproduce the observed abundances of COMs toward pre-stellar cores. All of them include grain-surface reactions and non-thermal desorption processes. One of these studies was carried out by \cite{vasyunin2017}, in which organic species are formed on dust grains through ion-molecule as well as radiative association reactions, and released in the gas phase through reactive desorption. With a multilayer treatment of ice chemistry and a special approach to reactive desorption based on  experimental studies of \cite{minissale}, the authors were able to reproduce the abundance of many COMs, like $\mathrm{CH_3OH}$ and $\mathrm{HCOOCH_3}$, toward the pre-stellar core L1544. Recent work by Shingledecker et al. (2018) has shown that taking into account cosmic-ray proton bombardment of icy dust grain mantles in gas-grain chemical models also boost the formation of some COMs. Despite the progress that has been made in the last few years, there are still some important pieces of the puzzle missing to understand the interstellar organic chemistry taking place in dark clouds.  

One of the missing pieces in oxygen-bearing chemistry has been ketenyl, $\mathrm{HCCO}$, a possible building block of oxygen-containing COMs. HCCO was detected for the first time in two sources, Lupus-1A and L483 by \cite{agundez}. After including all known formation and destruction reactions of HCCO from the literature and the chemical databases, it was found that the main formation mechanism for HCCO is the reaction between OH and $\mathrm{C_2H}$. Nevertheless, a discrepancy of three orders of magnitude between the observed and the predicted HCCO abundance implies that there is an effective formation mechanism of HCCO still unknown and which can compete against its depletion by reactions with neutral atoms \citep{agundez}.  
Previous studies \citep{hudson,maity} had shown that ketene, $\mathrm{H_2CCO}$, can be formed in ices after UV and electron irradiation, representing a possible formation path for HCCO as an intermediate product. 
In \cite{wakelam}, the rate coefficient of the main HCCO formation pathway was increased by almost an order of magnitude based on the comparison to similar reactions. In doing so, their chemical model was able to reproduce the observed abundances of HCCO towards the sources Lupus-1A and L483. However,  the reaction of interest, $\mathrm{OH + C_2H \rightarrow HCCO + H}$, has never been studied before, neither theoretically nor experimentally, and there remains a large uncertainty behind HCCO chemistry.
From the above considerations, it is evident that further observations are needed to revise the gas-grain chemistry of HCCO.
Moreover, deuterated ketenyl, DCCO, has never been detected in the ISM and its spectral line catalog is not yet available in the online databases. The interstellar detection of DCCO would give information about possible deuteration routes of HCCO, and put more constraints on current gas-grain chemical models. 

The laboratory spectrum of HCCO was first measured  by photoionisation mass spectroscopy \citep{jones} targeting the reactions
\begin{eqnarray}
\mathrm{C_2H_2} + \mathrm{O} \rightarrow \mathrm{HCCO + H} \\
                     \rightarrow \mathrm{CH_2 +CO}.
\end{eqnarray} 
where channel (1) is more than two times faster than reaction (2) throughout a broad range of temperatures \citep{michael}. Follow-up work was done by \cite{endo}, where pure rotational lines of HCCO ($N=15-18$) and DCCO  ($N=17-20$) were measured in the 320-390 GHz range. Their analysis showed a considerable interaction between the ground state and a low-lying excited state (Renner-Teller pair), which leads to a strong perturbation of the spin-rotation splitting. In another work done by \cite{ohshima}, HCCO was studied in the microwave spectral region in a supersonic jet expansion. In particular, the fine and hyperfine components of the $N=1-0$ transition were measured at around 21 GHz, enabling the determination of the hyperfine coupling constants ascribed to the hydrogen nucleus. A further study by \cite{szalay} involved the ab initio calculations of rotational constants, equilibrium geometries and excitation energies of the ground and excited electronic states of HCCO. Their results are in good agreement with the values obtained from the microwave and millimeter measurements.   

In this work we have recorded the pure rotational spectra of HCCO and DCCO between 170 and 650 GHz, extending substantially the frequency range compared to the previous studies mentioned above. More than 100 lines were recorded and assigned for each isotopologue, reducing the errors of the rotational and centrifugal distortion constants down by a factor of $\sim40$. With this new data set we are able to provide highly accurate rest-frequencies for future astronomical millimeter observations.

\section{Laboratory Experiments}
In the present work, the rotational spectra were taken with the frequency-modulation mm/sub-mm Absorption Cell spectrometer developed at the Center for Astrochemical Studies (CASAC) in Garching. A full description of the experiment can be found in \cite{bizz}. This instrument consists of an absoprtion cell made of a Pyrex tube (3 m long and 5 cm in diameter), which  serves as the main flow cell. Each end is equipped with one stainless steel electrode in order to apply a high voltage and ignite a discharge, creating a plasma from the appropriate mixture of gaseous precursors. To stabilize the plasma and minimize the collisions between the species, the outer walls of the cell are cooled with liquid nitrogen.

As a radiation source we use an active multiplier chain (Virginia Diodes, Inc.) which is coupled to a synthesizer (Keysight E8257D) operating at centimeter wavelengths. By applying additional frequency multipliers, the radiation source can cover continuously the 82-1100 GHz frequency range. High phase accuracy is achieved by phase locking the synthesizer with an external 10 MHz rubidium clock (Stanford Research Systems).  For the detection of the molecular signal we use a liquid-He-cooled InSb hot electron bolometer (QMC Instruments Ltd.). The signal is frequency-modulated with a sine wave of 15 kHz, while the output detector signal is demodulated at $2f$ (where $f$ denotes the modulation frequency) by a lock-in amplifier (Standford Research Systems SR830), meaning that we record the second derivative of the actual absorption line. 

HCCO and DCCO were produced by a DC high-voltage plasma ($\sim11 \, \mathrm{mA}$, $\sim0.6 \, \mathrm{kV}$) of a 2:1 mixture of $\mathrm{O_2}$ and $\mathrm{C_2H_2}$/$\mathrm{C_2D_2}$ diluted in a buffer gas of Ar (total pressure $\sim28 \, \mathrm{mTorr}$). The freezing point of acetylene $\mathrm{C_2H_2}$ provides a lower limit of the cooling temperature of the cell at around 190 \nolinebreak K.

\section{Results and data analysis}
\begin{figure*}[h]
	\centering
	\includegraphics[width = 1.0\textwidth]{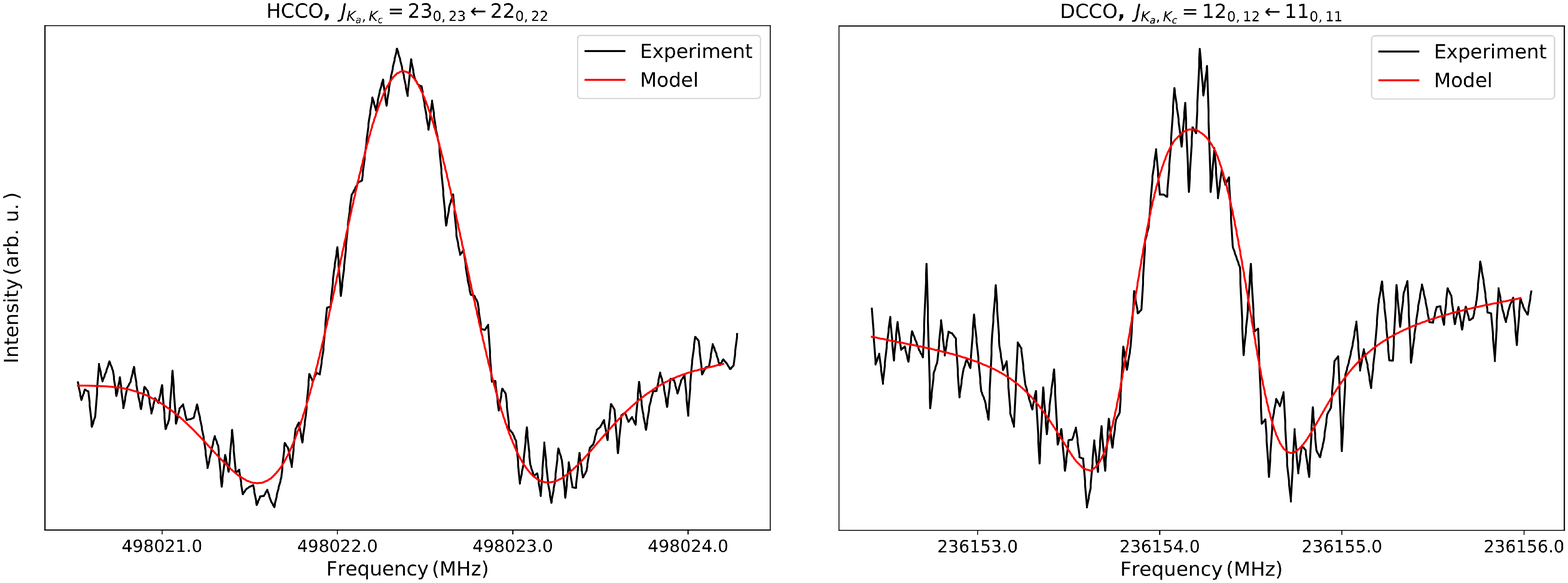}
	\caption{(\textit{Left panel}) Recording of the $J_{K_a, K_c} = 23_{0,23}-22_{0,22}$ transition of HCCO. Total integration time is 309 s with a time constant of 3 ms. (\textit{Right panel}) Recording of the $J_{K_a, K_c} = 12_{0,12}-11_{0,11}$ transition of DCCO. Total integration time is 127 s with a 3 ms time constant. The red curves represent our modeled $2f$ Voigt profiles (see text). }
	\label{Fig:spectrum}
\end{figure*}

HCCO and DCCO have a $^{2}A \arcsec$ ground electronic state with one unpaired electron at the carbon nucleus, showing paramagnetic nature \footnote{The ketenyl radical is a quasilinear molecule in the ground state with an $\mathrm{\angle HCC}$ angle of $135^\circ$ and a $\mathrm{\angle CCO}$ angle of $169^\circ$ \citep{szalay}.}. Both species show a rich rotational spectrum with fine (spin-rotation interaction) and hyperfine structure (H/D nuclear spin). In addition, the ground state rotational constants determined in \cite{endo} led to inertia defects of 0.155(15) and 0.120(4) $\mathrm{amu \, \mathring{A}}$ for HCCO and DCCO, respectively, confirming the planarity of the species. 
The dipole moment along the $a$ axis has been determined by \cite{szalay}  to be 1.59 D and by \cite{jerosimic} to be 1.68 D. We have adopted the second value for the predictions of the line intensities. The total dipole moment is only marginally larger than its $a$ component. We detected 114 and 138 new lines for HCCO and DCCO, respectively. These are $a$-type rotational transitions, with $N$ ranging from 8 to 27 for HCCO and with $N=9-33$ for DCCO with a maximum $K_a$ of 4.  We have modeled the frequency-modulated absorption lines using an in-house code that implements the $2f$ Voigt profiles presented by \cite{dore}. The estimated accuracy on the central frequency depends on the signal-to-noise ratio, the line width and the baseline (mainly caused by standing waves formed between the two windows placed at each end of the absorption cell), and ranges between 25-75 kHz by our estimates.  In Figure \ref{Fig:spectrum} we show plots of HCCO and DCCO lines, as well as the corresponding fitted profiles.

\begin{table*} 
\centering
\caption{Spectroscopic parameters determined for HCCO and DCCO$^\mathrm{a}$.} 
\label{tab:spec_parameters}
\setlength{\tabcolsep}{10pt}
\begin{tabular}{lr@{.}lr@{.}lr@{.}lr@{.}l}
\hline\hline \\[0.5ex]
Constants (MHz) &  \multicolumn{2}{c}{HCCO$^{\mathrm{b}}$} & \multicolumn{2}{c}{HCCO$^{\mathrm{c}}$} & \multicolumn{2}{c}{DCCO$^{\mathrm{b}}$} & \multicolumn{2}{c}{DCCO$^{\mathrm{c}}$} \\
\hline \\ [0.5ex]
$A$ &  1046955&(1682)  & 1243000&(45000) & 652933&(410) & 652100&(3600) \\ [1ex]
$B$ & 10896&7954(16) & 10896&788(41) & 9926&7979(12)  & 9926&8008(104) \\ [1ex]
$C$ & 10766&0154(23) & 10766&466(39) & 9755&2295(11) & 9755&2316(126)\\ [1ex]
$D_N$ & 0&0038207(13) & 0&003861(21)& 0&0035141(15) & 0&0035088(34)\\ [1ex]
$D_{NK}$ &  -0&2536(24) & 0&2376(26) & -1&67334(82) & -1&6724(123) \\ [1ex]
$D_K$ & 30475&(528) & 18480&(1200) & 3952&(264) & 5000&\\ [1ex]
$d_1$ &  -0&00008151(65) & -0&000119(34) & -0&00024091(40) & -0&0002431(73) \\ [1ex]
$d_2$ &  -0&00004013(90)& 0&0000201(101) & -0&00005828(90) & -0&0000649(43)\\ [1ex]
$H_{KN}$ & -0&191900(924) & -0&01185(24) & 0&037735(302) & 0&0375(45)\\ [1ex]
$H_K$ &  352&(14) & \multicolumn{2}{c}{-} & \multicolumn{2}{c}{-} & \multicolumn{2}{c}{-}    \\ [1ex]
$L_{KKN}$ &  0&020538(108) &  \multicolumn{2}{c}{-}  & -0&0106543(349) & -0&01059(52) \\ [1ex]
$P_{KKKN}$ & -0&00069701(376) &  \multicolumn{2}{c}{-}  & 0&00039998(121) & 0&0003971(176)\\ [1ex]
$\epsilon_{aa}$ & -247917&6(69) & -247827&(74) & -112219&2(33) & -112201&(34) \\ [1ex]
$\epsilon_{bb}$ &  -43&128(60) & -43&1(27) & -30&655(74) & -30&62(85)\\ [1ex]
$\epsilon_{cc}$  & 13&322(60) &13&8(27) & 10&225(65) & 10&20(69) \\ [1ex]
$(\epsilon_{ab} + \epsilon_{ba})/2$ &  529&4(77)  & 1619&(53) & 276&28(99) & 275&(31)\\ [1ex]
$D^{S}_N$ & -0&000086(60)  & -0&00920(178) & 0&000519(31) & 0&00041(65) \\ [1ex]
$D^{S}_{NK}$ & 2&405(36)  & 1&52(36) & -1&698(18) &-1&589(191) \\ [1ex]
$D^{S}_{KN}$ & -17&66(31) & -4&68(141) & 5&180(76) & 4&71(65) \\ [1ex]
$D^{S}_{K}$ &   -53&61(66) & 953&1(53) & 1&75(34) & 826&7(21) \\ [1ex]
$d^{S}_1$ & -0&000056(44) & -0&00520(141) &  0&000316(23) & 0&00025(44)\\ [1ex]
$H^{S}_K$ &  -23&320(60) & -27&386(169) & -21&6652(164) & -21&628(89) \\ [1ex]
$H^{S}_{KNK}$ &  0&1425(41) &  \multicolumn{2}{c}{-}   &  \multicolumn{2}{c}{-}  &  \multicolumn{2}{c}{-}   \\ [1ex]
$b_F$ & -54&0304(46) & -54&030(24)$^{\mathrm{d}}$ &  \multicolumn{2}{c}{-}  &  \multicolumn{2}{c}{-}  \\ [1ex]
$c$ & 16&029(18) & 16&041(92)$^{\mathrm{d}}$ &  \multicolumn{2}{c}{-}   &   \multicolumn{2}{c}{-} \\ [1ex]
$\sigma$  &  0&059 & 0&125 & 0&064 & 0&029  \\ [1ex]
$\sigma_w$  & 1&15 & \multicolumn{2}{c}{-} & 1&18 &  \multicolumn{2}{c}{-}   \\ [1ex]
No. of lines &  \multicolumn{2}{c}{168} & \multicolumn{2}{c}{54} & \multicolumn{2}{c}{193} & \multicolumn{2}{c}{55} \\ [1ex]
\hline \\
\end{tabular}
\tablefoot{${^\mathrm{a}}$ Values in parentheses are 1$\sigma$ uncertainties, expressed in units of the last quoted digit. $\sigma_w$ denotes the weighted root mean square deviation.} \\
\tablebib{${^\mathrm{b}}$ This work ; $^{\mathrm{c}}$  \cite{endo}; $^{\mathrm{d}}$  \cite{ohshima}}
\end{table*}

\begin{table*}
\centering
\caption{Spin-rotation constants $\epsilon_{aa}(K)$ determined for HCCO and DCCO. }
\label{table:spearman}
\setlength{\tabcolsep}{10pt}
\begin{tabular}{lr@{.}lr@{.}l}
\hline\hline \\[0.5ex]
Constants (MHz) &\multicolumn{2}{c}{HCCO}& \multicolumn{2}{c}{DCCO}\\
\hline \\[0.5ex]
$\epsilon_{aa}(K=0)$ & -247917&6(69)  & -112219&2(33) \\ 
$\epsilon_{aa}(K=1)$ & -157486&3(22) & -89141&7(18) \\
$\epsilon_{aa}(K=2)^\mathrm{a}$ &  -112607&9(31) & -71781&2(21)\\
$\epsilon_{aa}(K=3)^\mathrm{a}$ &  -83405&9(23) & -56987&9(16) \\
$\epsilon_{aa}(K=4)^\mathrm{a}$& -60804&3(16) & -43191&0(12) \\
\hline
\end{tabular}
\tablefoot{${^\mathrm{a}}$ Computed from the $\epsilon_{aa}(K=0)$ value. See text for explanation.}
\end{table*}

The initial prediction for new rotational transitions of HCCO and DCCO was done based on the spectroscopic constants determined previously by \cite{endo}. 
The spectral analysis was done using the Pickett SPFIT/SPCAT suit of programs \citep{pickett}, applying a S-reduced, asymmetric rotor Hamiltonian \citep{watson}:
\begin{eqnarray}
\mathrm{H = H_{rot} +H_{sr} + H_{hfs}},
\end{eqnarray}
where $H_{\mathrm{rot}}$ describes the rotational energy of the molecule, $H_{\mathrm{sr}}$ the spin-rotation interaction and $H_{\mathrm{hfs}}$ the hyperfine coupling. 
The spin-rotation interaction arises from the spin dipole moment of the unpaired electron that interacts with the magnetic field generated by the rotation of the molecule. The hyperfine coupling was only included for HCCO, since its hyperfine splitting was already resolved in \cite{ohshima}, unlike DCCO.
The hyperfine coupling involves two constants: the Fermi contact interaction $b_{F}$ and the dipolar interaction $c$. The Fermi contact interaction describes the magnetic coupling between the unpaired electron and the hydrogen nucleus, indicating the overlap of the unpaired spin on the nucleus. The dipolar coupling depends only on the orientation of the unpaired electron with respect to the hydrogen nucleus, meaning that constant $c$ is structure dependent \citep{ohshima}.

The previous study by \cite{endo} had shown that the observed spin splittings of the $K_a=2$ and $K_a=3$ lines are much smaller than those expected from the $K_a=1$ lines. This peculiar behaviour is caused by the Renner-Teller coupling (see Appendix \ref{renner}), whereby a special case of vibronic coupling is cancelling the degeneracy of the electronic state $^{2}\Pi_{u}$, which subsequently splits to an excited, linear component, $1 ^{2}A'(1 ^{2}\Pi)$, and a ground, bent component, $1 ^{2}A \arcsec(1 ^{2}\Pi)$. The low lying excited state interacts with the ground state, which leads to a non-negligible perturbation of the rotational levels.  As a result,  a conventional power series of the spin-rotation constant $\epsilon_{aa}(K)$ is not sufficient to reproduce the anomalous K-dependence. In order to include this distortion in the analysis, \cite{endo} adopted a Pade-type formula given by:
\begin{equation} \label{Eq:epsilon}
\epsilon_{aa}(K) = \epsilon_{aa}(K=0)/(1 + t \cdot K),
\end{equation}
allowing the prediction and assignment of transitions with $K_a$ up to 4. The parameter $t$ corresponds to $\omega/\Delta E$, where $\omega$ describes the bending vibration frequency and $\Delta E$ the electronic energy of the excited state (see work from \cite{endo} for further explanation). 
In order to take into account the energy perturbation mentioned above, we calculated one effective spin-rotation constant $\epsilon_{aa}(K)$ for every $K$ transition according to Equation \ref{Eq:epsilon}, using the $t$ value inferred by \cite{endo}. Thus, we defined five separate $\epsilon_{aa}$ constants in the input file of the fitting program, where $\epsilon_{aa}(K=0)$ is applied to all transitions with $K_a=0$, $\epsilon_{aa}(K=1)$ is applied to all transitions with $K_a=1$, and so on. 
After including all the lines, we were able to fit  $\epsilon_{aa}(K=0)$ and  $\epsilon_{aa}(K=1)$ while keeping the remaining epsilon parameters fixed. We also set the relation between $\epsilon_{aa}(K=0)$ and the other epsilon parameters for $K=2,3,4$ constant. This means that $\epsilon_{aa}(K=2,3,4)$ are changed with respect to the fitted $\epsilon_{aa}(K=0)$ parameter, following Equation \ref{Eq:epsilon}. 
The application of the Pade-type expansion for $\epsilon_{aa}(K)$ allows us to fit and predict rest frequencies for transitions with $K_a$ up to 4 and an accuracy well suited for astrophysical needs. Without this approach we were only able to predict lines up to $K_a=2$.

The final fits include our new measured lines as well as those measured in the previous studies mentioned above.
In case of HCCO, there were two transitions, $J_{K_a, K_c} = 15_{3,12}-14_{3,11}$ and $J_{K_a, K_c} = 16_{3,13}-15_{3,12}$, measured by \cite{endo}, which were considerably deviating from the predicted frequencies ($\sim300$ kHz). For this reason, we have re-measured these two lines in the laboratory to search for possible misassignments.
However, since the discharge of acetylene and oxygen leads to the production of numerous species (other than HCCO), the resulting spectra had various lines, that were partially blended with the HCCO features, making the line assignment rather difficult. In the end, we were able to measure and assign the $J_{K_a, K_c} = 16_{3,13}-15_{3,12}$ transition, which is in better agreement with the prediction than the one measured by \cite{endo}.

For both species, the final fits allowed us to determine all the quartic centrifugal distortion constants,  one sextic parameter, $H_{KN}$, as well as all the quartic and one sextic spin-rotation constant, $H^{S}_K$ \footnote{The order of the rotational constant refers to which power of $J$ the energy contributions depend on.}. The resulting parameters are listed in Table 1. For the HCCO spectral analysis we additionally included one sextic constant $H_K$, one octic, $L_{KKN}$, and one decadic constant, $P_{KKKN}$, along with one sextic spin-rotation parameter, $H^{S}_{KNK}$, compared to the previous fit done by \cite{endo}. Adding these extra parameters has considerably improved the fit, resulting in a root mean square (rms) deviation of 59 kHz, which is an improvement of a factor $\sim2$ compared to the previous fit by \cite{endo}  (rms =125 kHz). 

Strong correlations between  $H_K, L_{KKN}$,  $P_{KKKN}$ and other constants modify the final values of some rotational parameters with respect to \cite{endo}. 
In particular, the introduction of $L_{KKN}$,  $P_{KKKN}$ has a significant impact on the constants  $d^{S}_1$, $(\epsilon_{ab} + \epsilon_{ba})/2$,   $D^{S}_N$, $H_{KN}$ and $D_{NK}$,  with correlation coefficients ranging from 82\% to 99\%. This also results in a sign change for $D_{NK}$, $d_2$ and $D^{S}_K$ compared to the previous published results \citep{endo}.
Nevertheless, these high-order centrifugal distortion corrections are critical to reproduce our spectral data within the experimental accuracy: by using a smaller set of constants the obtained rms is increased up to several MHz.  An accurate determination of the rotational constants would require a full analysis of both interacting states ($1 ^{2}A'(1 ^{2}\Pi)$ and  $1 ^{2}A\arcsec(1 ^{2}\Pi)$), which is beyond the scope of this work.

The spectral analysis of DCCO resulted to a root mean square deviation of 64 kHz. Also here the rotational parameters are more strongly constrained in our fit (see for example constants $(\epsilon_{ab} + \epsilon_{ba})/2$ and $D^{S}_N$) compared to \cite{endo}. In addition, we were able to determine $D_K$, unlike \cite{endo}, where the constant had previously been kept fixed to 5000 MHz.  In case of $D^{S}_K$  we note a considerable difference of a factor of $\sim 500$ between our and the previous study.  Since there is a correlation between the constants $D_{K}$ and $D^{S}_K$ of $\sim 50\%$, we do expect a change in $D^{S}_K$ by fitting $D_K$.

Table 2 shows a summary of the $\epsilon_{aa}(K)$ constants for HCCO and DCCO determined in this work.  A list with all the experimental frequencies is available at the CDS as supplementary data.

\section{Astrochemical Relevance}

In cold environments with a gas temperature of 10 K or lower (such as Lupus-1A and L483, where HCCO has been detected), the Boltzmann distribution for HCCO and DCCO peaks at around 100 GHz. With the present results we are able to reduce the predicted uncertainties down to 5-14 kHz at this frequency range (for HCCO and DCCO, respectively), opening the possibility for astronomical observations towards cold sources, such as starless and pre-stellar cores, which are characterized by narrow emission lines. Table \ref{tab:spec_parameters} lists five DCCO transitions with $K_a=0$ that could be relevant for observations of cold sources in the 3 mm range. 

\begin{figure*}[h]
	\centering
	\includegraphics[width = 1.0\textwidth]{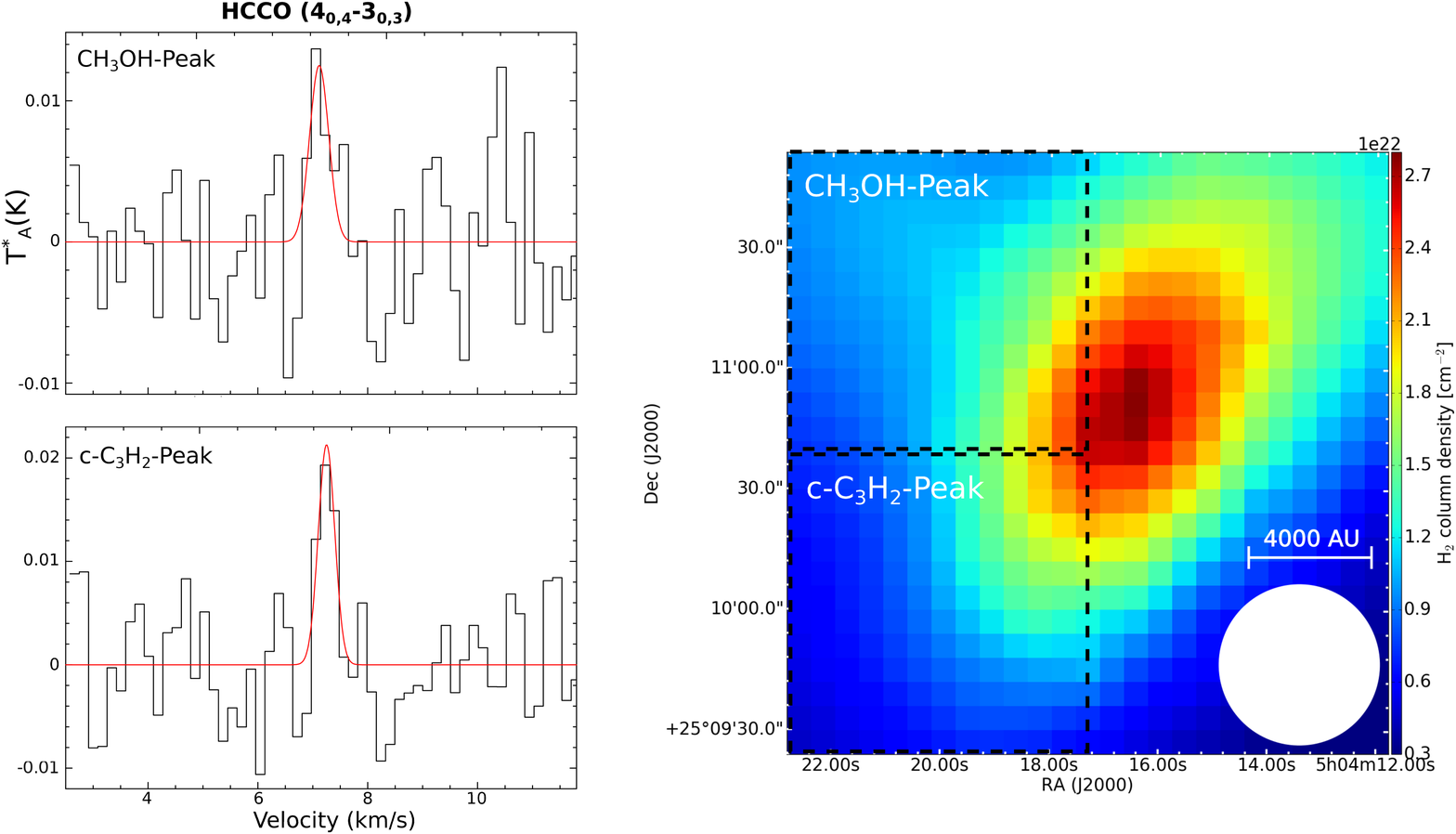}
	\caption{Map of the $\mathrm{H_2}$ column density derived from far-infrared images taken by $\textit{Herschel}$ \citep{spezzano16}.  The dashed squares cover 40$\arcsec$ in the x- and y- direction and indicate the regions where the molecules $c$-$\mathrm{C_3H_2}$ and $\mathrm{CH_3OH}$ peak. The white circle represents the beam of the \textit{Herschel}/SPIRE instrument. On the left part of the figure we show the resulting spectra of the $J_{K_a, K_c} = 4_{0,4}-3_{0,3}$ transition of HCCO averaged over the dashed squares. The red line plots the CLASS Gaussian fit.}
	\label{Fig:L1544_emission}
\end{figure*}

\begin{table*} [h]
\centering
\caption{Spectroscopic parameters of DCCO transitions relevant at low temperatures ($\sim10$ K).}
\label{tab:spec_parameters}
\setlength{\tabcolsep}{10pt}
\begin{tabular} {c c c c c c} 
\hline \hline \\
Transitions & Frequency &  Accuracy  &  $E_{\mathrm{up}}$  & $A_{ul}$ & $g_u$ \\
$\mathrm{J}_{K_{a},K_{c}}$ & (MHz) & (kHz) &  ($ \mathrm{K}$) & ($10^{-5} \mathrm{s^{-1}}$)  &  \\
\hline \\ [-1ex]
$4_{0,4}-3_{0,3}$  & 78721.6685 & 14.4 & 9.45 & 0.71  & 10 \\
$5_{0,5}-4_{0,4}$  & 98402.4782 & 14.4 & 14.18 & 1.42 & 12 \\
$6_{0,6}-5_{0,5}$  & 118082.6947 & 14.4 & 19.85 & 2.50 & 14  \\
$7_{0,7}-6_{0,6}$  & 137762.1984 & 14.4 & 26.46 &  4.01 & 16  \\
$8_{0,8}-7_{0,7}$  & 157440.8700 & 14.4 & 34.03 & 6.03 & 18  \\
\hline
\end{tabular} 
\tablefoot{$E_{\mathrm{up}}$ is the upper state energy, $A_{ul}$ is the Einstein coefficient of the corresponding transition and the upper state degeneracy is given by $g_u$. }
\end{table*}

Further HCCO detections towards cold sources will help us clarify the gas-grain chemistry of this species in the early phase of star formation. As already mentioned in the introduction, HCCO is believed to be formed mainly in the gas-phase, via the reaction of $\mathrm{C_2H}$ with OH. A first step towards understanding the HCCO interstellar chemistry would be to search for a correlation between HCCO and its main precursor $\mathrm{C_2H}$. An ideal candidate for this purpose is the well-studied, evolved pre-stellar core, L1544.
The chemical structure of the source L1544 was studied in \cite{spezzano17} through molecular maps. In particular, the authors investigated the spatial distribution of 22 different molecules, and came to the conclusion that different parts of the source favour the formation of either carbon-, nitrogen or oxygen-bearing molecules.  $c$-$\mathrm{C_3H_2}$ and other carbon-chain molecules like $\mathrm{H_2CCC}$ and $\mathrm{C_3H}$, have their emission peak in the south-east part of the core, while oxygen-bearing species, such as $\mathrm{CH_3OH}$, HCO and SO peak towards the north-east part of the dust continuum peak.  In addition, HNCO (as well as $\mathrm{CH_3CCH}$, $\mathrm{CH_2DCCH}$ and $\mathrm{CH_3CCD}$) peaks towards the north-west part of the core. 
The right part of Figure \ref{Fig:L1544_emission} shows the $\mathrm{H_2}$ column density map in L1544, which was derived by the far-infrared images taken with \textit{Herschel}/SPIRE \citep{spezzano16}. 
We used the data by \cite{spezzano17} to search for an HCCO line at 86.6 GHz. For the data processing we employed the software CLASS from the GILDAS\footnote{http://www.iram.fr/IRAMFR/GILDAS} package \citep{pety}. Since the HCCO detections are very weak ($\sim 10$ mK), we chose to average its spectrum over two large regions that contain either the $c$-$\mathrm{C_3H_2}$ or $\mathrm{CH_3OH}$ peak, and cover $40\arcsec$ in the x- and y-direction, respectively. These regions are marked by the dashed squares in Figure \ref{Fig:L1544_emission}. The center of the map is positioned at the 1.3 mm dust peak \citep{ward}. 

As it can be seen on the left part of Figure \ref{Fig:L1544_emission}, HCCO peaks towards the region where $c$-$\mathrm{C_3H_2}$ peaks, while the line intensity decreases by almost a factor of 2 towards the $\mathrm{CH_3OH}$ peak.This is a clear evidence that HCCO follows a different chemistry than other oxygen-bearing species, showing instead a predominant hydrocarbon chemistry.  This  favours the fact that $\mathrm{C_2H}$ is chemically related to HCCO. However, higher-sensitivity observations are needed to prove this point and constrain the gas-grain chemical models.

\section{Chemical Model}
The chemical model that we apply is based on the model developed by \cite{vasyunin2013} with several minor updates in grain-surface chemistry (Vasyunin et al. 2019, in prep.), and includes a gas-grain chemical network with 6000 gas-phase reactions, 200 surface reactions and 660 species. The gas-phase and grain surface chemistry are connected by accretion and desorption processes. The code numerically integrates chemical rate equations and gives a set of time dependent abundances of all chemical species for a time span that the user chooses. 
We use typical conditions of a cold dark cloud with a gas and dust temperature of 10 \nolinebreak K, a visual extinction of 30 magnitudes, a proton density of $2 \cdot 10^{4} \, \mathrm{cm^{-3}}$ and a cosmic ionization rate of $1.3 \cdot 10^{-17} \, \mathrm{s^{-1}}$.
As initial atomic abundances we use the ones reported in \cite{wakelam}.
In addition, we include all the known formation and destruction reactions for HCCO available in the literature and the KIDA (KInetic Database for Astrochemistry) online database \citep{wakelam2}. In order to simulate the HCCO chemistry, the model is run with steady physical conditions and integrated over $10^7$ years. As already reported in previous studies \citep{agundez, wakelam}, we find that the main formation pathway of HCCO in the gas-phase is the reaction of $\mathrm{C_2H}$ with OH, while the most effective reaction on grain surfaces is $\mathrm{s-H + s-CCO \rightarrow s-HCCO}$. For the reaction $\mathrm{OH + C_2H \rightarrow HCCO + H}$ we use a  rate coefficient of $2\cdot10^{-10} \, \mathrm{cm^3 \, s^{-1}}$, as proposed in \cite{wakelam}. 

In the end, our simulations are able to reproduce the observed abundance of HCCO for times between $10^5$ and $10^6$ yr, reaching a maximum abundance of $5.6\cdot10^{-11}$.  
Nevertheless, a theoretical and/or an experimental study of the main formation pathway $\mathrm{OH + C_2H \rightarrow HCCO + H}$ is still needed for a better interpretation of the chemical simulations.

\section{Conclusions}
This work describes a rotational analysis of the ketenyl radical, HCCO, and its deuterated counterpart DCCO. For each isotopologue we measured and assigned more than 100 lines. We extended the measurements up to $\sim650$ GHz, which helped us improve the determination of the spectroscopic parameters with respect to the previous study by \cite{endo}. In particular, we measured the rotational transitions with $N$ ranging from 8 to 27 for HCCO and with $N=9-33$ for DCCO with a maximum $K_a$ of 4. 

\begin{itemize}
\item For both species we were able to determine all the quartic centrifugal distortion constants,  one sextic parameter, $H_{KN}$, as well as all the quartic and one sextic spin-rotation constant, $H^{S}_K$. \\

\item We expanded the centrifugal distortion analysis of HCCO by including a sextic constant $H_K$, an octic, $L_{KKN}$, and one decadic constant, $P_{KKKN}$, along with one sextic spin-rotation parameter, $H^{S}_{KNK}$. This extended set of parameters reduced the rms deviation by a factor of $\sim2$ with respect to the previous fit by \cite{endo}. We also decreased the uncertainty on the rotational constants up to a factor of $\sim40$. \\

\item We show that towards the pre-stellar core L1544, HCCO peaks within the region where $c$-$\mathrm{C_3H_2}$ peaks, suggesting that the ketenyl formation is  based predominantly on hydrocarbon chemistry.\\

 \item The spectral analysis of DCCO resulted to a rms deviation of 64 kHz. The rotational parameters are more strongly constrained in our fit with respect to \cite{endo}. We provide for the first time a catalog of highly accurate frequencies for DCCO (uncertainties at 3 mm $\sim 15$ kHz), which allows a future DCCO detection in cold sources, like starless and pre-stellar cores. 
\end{itemize}

\begin{acknowledgements}
We thank the anonymous referee for his/her comments that
significantly improved the present manuscript. The work by Anton Vasyunin is partially supported by the ERDF project ``Physical and chemical processes in the interstellar medium'', No.1.1.1.1/16/A/213.
\end{acknowledgements}

\newpage 

\appendix

\section{Theoretical considerations}

\subsection{Renner-Teller Effect} \label{renner}
The coupling between the vibrational angular and the electron orbital momenta (with $\Lambda > 0$) in a linear molecule is known as the Renner-Teller interaction \citep{herzberg, renner}. This effect occurs when the nuclei are displaced within a bending vibration and a vibrational angular momentum with the projection number $l=0,1,2,..$ is induced. The resulting vibronic angular momentum $K$ around the molecular axis is then equal to $K =  |\pm \Lambda \pm l\,|$, meaning that a doubly degenerate electronic state (with $\Lambda > 0$) is split into two components \citep{brown,lee}. In the case of the ketenyl radical, the unpaired electron occupies a $\mathrm{\Pi}$ orbital that lies perpendicular to the molecular plane. When the H-atom is bent away from the linear molecular axis,  the $1 ^{2}\Pi_{u}$ degeneracy is cancelled and split into two components: a $1 ^{2}A \arcsec(1 ^{2}\Pi)$ bent component described by an attractive potential with respect to the bending coordinate, and a $1 ^{2}A' (1 ^{2}\Pi)$ linear component described by a repulsive potential \citep{szalay, aarts}. The anomalous features observed in the spectrum are ascribed to the fact that $^{2}A'$ is a low-lying excited state that interacts with the ground state $^{2}A\arcsec$ resulting to a non-negligible perturbation of the rotational levels. A small energy splitting between these two states would mean a stronger interaction of the systems, and thus a stronger perturbation. 
Using the spin-rotation coupling obtained from the submillimeter spectrum, \cite{endo} 
predicted that the excited electronic state $1 ^{2}A'$ lies only 777 K above the ground state $ 1 ^{2}A \arcsec$, assuming that the unpaired electron is mostly localized at the carbon atom. However, \cite{szalay1992} showed that the unpaired electron has a considerable density on both the oxygen and the carbon atom, which led finally to an energy splitting of 1727 K \citep{szalay}.

\end{document}